\documentstyle[eqsecnum,epsf,aps,prb,twocolumn]{revtex}
\newcommand{\eq}{\begin{equation}}
\newcommand{\ee}{\end{equation}}
\newcommand{\eqa}{\begin{eqnarray}}
\newcommand{\eea}{\end{eqnarray}}

\def\zt{{z_T}}
\newcommand{\pprl}{Phys. Rev. Lett.\ } 
\newcommand{\pprb}{Phys. Rev. {B}}

\def\calg{{\cal G}}
\def\calgr{{\cal G}_{\rm reg}}
\def\cala{{\cal A}}

\def\tphi{{\tau_\varphi}}
\def\tee{{\tau_{\rm ee}}}

\def\lphi{{L_\varphi}}

\begin{document}
\twocolumn[
\hsize\textwidth\columnwidth\hsize\csname@twocolumnfalse\endcsname
\draft
 
\title{Short-Range Interactions and Scaling Near Integer Quantum Hall Transitions}
\author{Ziqiang Wang$^a$, Matthew P. A. Fisher$^b$, S.~M. Girvin$^c$, 
J.~T. Chalker$^d$}
\address{$^a$ Department of Physics, Boston College, Chestnut Hill, MA 02167\\
$^b$ Institute for Theoretical Physics, University of California,
Santa Barbara, CA 93106-4030\\
$^c$ Department of Physics, Indiana University, Bloomington, IN 47405\\
$^d$ Theoretical Physics, Oxford University, Oxford OX1 3NP, United Kingdom}
\date{\today}
\maketitle

\begin{abstract}
We study the influence of short-range electron-electron interactions on 
scaling behavior near the integer quantum Hall plateau transitions. 
Short-range interactions are known to be irrelevant at the renormalization 
group fixed point which represents
the transition in the non-interacting system. We find, nevertheless, that 
transport properties change discontinuously when interactions are introduced. 
Most importantly, in the thermodynamic limit the conductivity
at finite temperature is zero without interactions,
but non-zero in the presence of arbitrarily 
weak interactions. In addition, scaling as a function of frequency, 
$\omega$, and temperature, $T$, is determined by the scaling variable 
$\omega/T^p$ (where $p$ is the exponent for the temperature dependence 
of the inelastic scattering rate) and not by $\omega/T$, as it would be at 
a conventional quantum phase transition described by an interacting 
fixed point. We express the inelastic exponent, $p$, 
and the thermal exponent, $z_T$, in terms of the scaling dimension, 
$-\alpha < 0$, of the interaction strength and the dynamical 
exponent $z$ (which has the value $z=2$), 
obtaining $p=1+2\alpha/z$ and $z_T=2/p$.
\end{abstract}
\pacs{PACS numbers: 73.40.Hm, 05.30.-d.}
]

\section{Introduction}
In this paper we study the effects of short-range interactions on the
nature of the transitions between quantized Hall plateaus in a
disordered two-dimensional electron gas (2DEG).\cite{qhebooks} These transitions are generally believed to be prime examples of continuous quantum
phase transitions, that is to say, examples of quantum critical
phenomena.\cite{hp,pruisken,steve,subirbook}
We focus here on
samples with sufficiently strong disorder that fractional quantum Hall
states do not intervene, so that the transitions are directly from one
integer Hall plateau to another. 
Recently, Shahar and collaborators have presented an analysis of
transport measurements that would seem to indicate an absence of a  
true quantum Hall liquid---insulator phase transition.\cite{shaharcaveat} 
The full implications of this are unclear at present,
but we presume that this is an indication of the
difficulty of reaching the asymptotic quantum critical regime 
in certain classes of disordered systems and will not consider it further
in this paper.

The existence of quantized Hall plateaus is intimately related to the
presence of disorder. In a single-particle description,
all states are localized except
for those at a single critical energy near the center of each Landau level.
Thus
the quantum phase transition is an unusual insulator to insulator
transition with no intervening metallic phase. The critical point itself
is quasimetallic, exhibiting anomalous diffusion.\cite{john} 
Associated with each transition between plateaus in $\sigma_{xy}$ there
is a peak in $\sigma_{xx}$ which in principle becomes infinitely sharp
at zero temperature (see however Ref.[\onlinecite{shaharcaveat}]) and
whose peak value is universal and close to\cite{ravin} $0.5 e^2/h$. 
However, as we discuss below, since we have the peculiar
circumstance that the set of extended states has measure zero, the zero
temperature limit is quite singular in the absence of interactions. In
the non-interacting case $\sigma_{xx}$ is actually rigorously zero in the limit of large sample size at
all values of the magnetic field, including the critical values, for any non-zero temperature. Moreover, it has been argued previously, using a combination of renormalization group techniques and numerical 
calculations\cite{lwinter} that interactions of sufficiently short range are
perturbatively irrelevant at the non-interacting fixed point. Hence
systems with short-range interactions scale into this singular non-interacting limit. We show in this paper that although interactions are irrelevant in this sense, they generate a non-zero critical value of $\sigma_{xx}$ and determine the nature of tempe
rature and frequency scaling near the critical point.
We expect that interactions have similar consequences near other delocalization transitions at which they are formally irrelevant, although behavior in a different category is possible if interactions are sufficiently strongly 
irrelevant. We note that irrelevant interactions which control dynamical
properties at a quantum critical point have been encountered
previously, in the theory of metallic spin glasses. \cite{sro}

In contrast to short-range, model interactions, true Coulomb interactions are believed to be relevant at
the non-interacting fixed point.\cite{lwinter} Hence one expects that
the true critical point is interacting. One of the persistent mysteries
in this problem is the fact that the experimentally observed value of
the correlation exponent $\nu$ at the interacting fixed point appears to
agree rather well with that predicted by numerical simulations of the
non-interacting fixed point.\cite{lwinter,allan}   That is, the
correlation length exponent does not appear to change even though the
value of the dynamical critical exponent $z$ is believed to change from $z=1$ for 
long-range interactions to $z=2$ for the short-range case.\cite{steve}
In the following, we do not consider this issue, and instead restrict our attention to short-range interactions.
The Coulomb interaction can be made short-range by placing a metallic
screening gate (ground plane) nearby. Such a situation was successfully
realized by Jiang, Dahm and collaborators\cite{hongwen} although they did not study the
quantum critical point, but rather the insulating phase at
densities well below the $0 \rightarrow 1$ plateau transition. They observed
that the variable range hopping exponent changed from the Efros--Shklovskii 
value expected for long-range interactions to the Mott value
expected for short-range interactions. 

The remainder of the paper is organized as follows. We summarize the scaling description of the quantum Hall plateau transitions in the next section, and discuss in section III the pathologies associated with the finite temperature
scaling behavior of the conductance in the non-interacting theory. From section
IV onwards, systems with short-range interactions are considered. We first describe
dephasing in the critical regime and the emergence of a long coherence
time, and determine the inelastic exponent $p$,
and the thermal exponent $\zt\neq z$ in terms of the scaling dimension of the
interactions. The difficulties arising from a direct application of conventional
scaling ideas are discussed. In section V, finite temperature
scaling is analyzed in the presence of short-range interactions. We
show that, although short-range interactions are formally irrelevant, they control aspects of the critical behavior. We demonstrate that the critical conductivity is non-zero provided interactions are not too strongly irrelevant. Finally, we construct new scaling variables and examine to what extent
conductance scaling can be forced into the conventional scaling framework.
Finite frequency scaling at $T=0$ is discussed in section VI and the
general scaling in temperature and frequency in section VII. Concluding remarks are presented in section VIII.

\section{Plateau Transitions and Scaling Theory}

The integer quantum Hall transition (IQHT) is driven by varying the location of the chemical potential, $\mu$, relative to the critical value, $\mu_c$. Throughout this paper we denote the distance from the critical point by $\delta = |\mu - \mu_c|$. Since
$\mu_c$ is dependent on magnetic field, $B$, the transition
is often reached experimentally by changing $B$ while keeping electron density fixed.
In the large $B$ limit, $\mu_c$ lies near the center of the Landau
levels. A body of experimental data, reviewed for example in Ref.\,[\onlinecite{steve}], can be summarized by the statements that: (i) On either side of the transition ($\delta\ne0$)
the Hall conductivity is quantized and the dissipative conductivity
has the limit $\sigma_{xx}\to0$ at zero temperature; (ii) At the transition ($\delta=0$) the Hall conductivity is unquantized and $\sigma_{xx}$
remains finite at zero temperature, so that the critical state is conducting.

Critical behavior is cut off in the presence of a finite length scale. In this event, the transition has a finite width $\delta^*$ within which the Hall conductivity deviates from
the quantized values and $\sigma_{xx}$ is non-zero. This width is 
\eq \frac{\delta^*}{\delta_0}\sim {\rm min}\left[\left(\frac{L_0}{L}\right)^{1/\nu}\!\!,\,
\left(\frac{T}{T_0}\right)^{1/\zt\nu}\!\!,\,\left(\frac{\omega}{\omega_0}\right)^{1/z\nu}\right] \label{dbscaling} \ee
where $L$, $T$
and $\omega$ are the finite system size, temperature and measurement
frequency in a specific experimental situation, and $\delta_0$, $L_0$, $T_0$ and $\omega_0$ are microscopic scales. The various exponents appearing in
Eq.~(\ref{dbscaling}) have the following meaning: $\nu$ is the exponent
of the single divergent length scale, the localization length $\xi\sim
\delta^{-\nu}$; $z$ is the dynamical exponent defining the length scale
introduced by a finite frequency, $L_\omega\sim \omega^{-1/z}$; and
$\zt$ is the thermal exponent governing a temperature-dependent length
scale $L_\varphi\sim T^{-1/\zt}$. In the conventional dynamical scaling
description of a quantum phase transition in which interactions are relevant and scale
to a finite strength at the transition, $\zt$ is expected
to be the same as $z$. All the three regimes in Eq.~(\ref{dbscaling})
have been probed experimentally,\cite{hp,koch,hpnew,lloyd} as well as
the regime in which electric field strength sets the cut-off.\cite{chow} Summarizing the results in the form in which they appear in the literature, we have
$\nu=2.3\pm0.1$, $1/\zt\nu=0.42\pm0.04$, and $1/z\nu=0.41\pm0.04$. This suggests
that $\zt=z=1$, which is consistent with the interpretation that the
Coulomb interaction is relevant at the transition. 
More generally, $\zt$ and $z$ may be independent exponents at a quantum phase transition. We show in the following that this is the case at the IQHT if the interaction scales to zero at the 
critical point. This happens for short-range interactions and could be realized
experimentally by screening out the long-ranged Coulomb interaction
with nearby ground planes or gates.

We now turn to recent theoretical developments. The Hamiltonian of interest 
describes interacting electrons moving in a two-dimensional
random potential in the presence of a magnetic field:
\eqa 
H&=&\sum_i\left[{1\over2m}\left({\vec
p}_i+{e\over c}{\vec A}\right)^2 +V_{\rm imp}({\vec r}_i)\right]
\nonumber \\ &+&{1\over2}
\sum_{i\neq j}V({\vec r}_i -{\vec r}_j), \label{h} \eea
where ${\vec A}$ is the external vector potential, $V_{\rm imp}$ is the
one-body impurity potential, and $V$ is the two-body interaction
potential. We write \eq V({\vec r}_i-{\vec r}_j)={u\over\vert {\vec
r}_i-{\vec r}_j \vert^\lambda}, \label{inter} \ee where $u$ and
$\lambda$ parameterize the strength and the range of the
interaction.\cite{notelambda} 
The existence of the IQHT in the model is not dependent on interactions, and the non-interacting theory, obtained by setting $u=0$, provides a simplified
but concrete model which has allowed extensive quantitative
calculations.\cite{bodo,sankar} A good understanding of the
main features of the non-interacting critical point has emerged: the static
localization length exponent has the value $\nu\approx 2.33\pm0.03$ and the dynamical
exponent is $z=d=2$. However, the relevance of the free electron model to the
IQHT in real materials depends on the nature and the effects of
electronic interactions. 

Imagine starting with a system at the noninteracting fixed point (NIFP),
and switching on the interaction. One can ask whether this interaction is a relevant or irrelevant perturbation
in the renormalization group (RG) sense. Such a stability analysis of
the NIFP has been performed.\cite{lwinter} For the unscreened Coulomb
interaction, $\lambda=1$ in Eq.~(\ref{inter}), $u$ has RG scaling dimension one
and is therefore a relevant perturbation.
The resulting flow away from the NIFP presumably leads to another, interacting fixed point (IFP) at which the effective interaction strength is finite. Critical phenomena in this case should be described by conventional
dynamical scaling theory with two independent critical exponents, $z$ and
$\nu$, and $\zt=z$. While one expects that $z=1$ on general 
grounds with Coulomb interactions,\cite{matthew} the value
of $\nu$ is unknown and may be different from the value at
the NIFP. Nevertheless, a scenario whereby Coulomb interaction changes
$z$ but not $\nu$ from the noninteracting values has been conjectured.
\cite{lwinter,wx} An alternative possibility \cite{ps,bodo3} is that there are
two divergent lengths at the critical point, with different exponents.

We shall not consider long-range Coulomb interactions further. Instead, we
focus on the case of short-range interactions having $\lambda>2$. As
mentioned above, this case is physically relevant when the IQHT is studied in the presence of ground planes or metallic gates. It has
been shown that for screened Coulomb interactions with
$\lambda>2+x_{4s}$, $x_{4s}\simeq0.65$, the RG dimension of $u$ is
$-\alpha=-x_{4s}$, so that interactions are an irrelevant
perturbation.\cite{lwinter} Notice that, in particular, the
dipole-dipole interaction has $\lambda=3$ and thus belongs to this class
of interactions. Moreover, for $x_{4s}>\lambda-2>0$, the interaction is
still irrelevant with the scaling dimension $-\alpha=2-\lambda$.
\cite{lwinter,bodointer}. In all these cases, the effective interaction
scales to zero at the transition in the asymptotic limit. The NIFP is
therefore stable against interactions. As a result, $\nu\approx 2.33$ and $z=2$. 
It turns out, though, that
short-range interactions, although irrelevant, control the
finite temperature behavior of the conductance. 
As we shall see, 
the scaling function for the conductance is discontinuous at 
zero interaction strength when written in terms of a natural set of 
scaling variables. We will show that the scaling theory thus
becomes unconventional, and a third independent critical exponent, the thermal
exponent, $\zt$, emerges in the scaling arguments. The value of $\zt$ is
set by the scaling dimension, $\alpha$, of the interaction strength: consideration of the dephasing time in the critical regime leads to $\zt=2z/(z+2\alpha)$. Since $\zt$
determines the transition width in the temperature scaling regime (cf
Eq.~(\ref{dbscaling})), experiments can, in principle, determine the
scaling exponent $\alpha$. We find, on the other hand, that the
frequency scaling of the conductance in this case is conventional, with
$z=d=2$, where $d$ is the spatial dimension of the system. We argue that
quantum critical scaling behavior of this kind may be a
general feature of finite temperature transport near 
quantum critical points, when interactions are irrelevant. The central feature is the
existence of a time scale, the dephasing time $\tau_\phi\sim T^{-p}$
where $p=1+2\alpha/z$, which is longer than the single characteristic
time, $\hbar/T$, at a conventional quantum phase transition. The long
coherence time results from the underlying free fermion description
and its associated infinite number of conservation laws. As a result, for
$\omega,T\neq0$, the $\omega/T$-scaling in conventional quantum phase
transitions\cite{steve} is replaced by
$\omega/T^p$-scaling.\cite{sondhi-kivelson} 

\section{Noninteracting Theory, $u=0$}
\subsection{T=0}
We begin by describing the finite size scaling of the zero frequency
conductance in the absence of interactions.\cite{sondhi-discussion}
Consider a 2D square sample of size $L\times L$. At $T=0$, 
the dimensionless conductance should depend only on $L/\xi$. 
Measuring the conductance in units of $e^2/h$, we write
\eq
g(\delta,L)=\calg_0(\delta L^{1/\nu}).
\label{gel}
\ee
The scaling function $\calg_0$ has the limiting behavior
\eq
\calg_0(X)=\left\{ \begin{array}{ll}
g_c, & X\to0, \\
0, & X\to\infty, \end{array} \right.
\label{gel0}
\ee
where $g_c$ is a critical conductance at the transition.
This quantity is expected to be universal for a given geometry
and boundary conditions.\cite{ravin,wjl,cho,wj}
In phase coherent, square samples under periodic transverse boundary conditions,
$g_c\simeq0.5$.
The behavior of $\calg_0(X)$ is known from numerical work in various
settings \cite{ando,gammel}, and in most detail for square samples from
transfer matrix calculations of the two-terminal Landauer conductance
\cite{cho,wj}: the results of these are sketched in Fig.~1a. 
It decays exponentially
for large $X$, according to $\calg_0(X)\sim\exp(- c X^\nu)$, where $c$ is a constant. Hence, in the limit $L\to\infty$,
$g$ is zero for all $\delta$ except $\delta=0$ at which it has
the finite value $g_c$, as shown in Fig.~1b.
We will denote the conductance in the thermodynamic limit, the quantity
of interest throughout the paper, by
suppressing the $L$ dependence in its argument. Thus
\eq
g(\delta)=\left\{ \begin{array}{ll}
g_c, & \delta=0, \\
0, & {\rm otherwise.} \end{array} \right.
\label{g0}
\ee
\begin{figure}   
\center   
\centerline{\epsfysize=1.8in   
\epsfbox{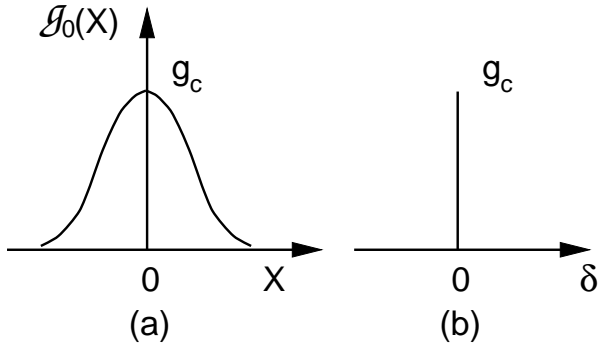}}   
\begin{minipage}[t]{8.1cm}    
\caption{(a) The conductance scaling function defined in Eq.~(\ref{gel0}), for the noninteracting theory.
(b) The behavior of the conductance in the noninteracting theory, in the thermodynamic limit
at zero temperature.
}  
\label{fig1}     
\end{minipage}    
\end{figure}   

\subsection{$T\ne0$}

For noninteracting electrons, the conductivity at $T\neq0$ is
\eq
\sigma_{xx}(\delta,T,L)=\int dE\left(-{\partial f\over\partial E}\right)
\calg_0(E L^{1/\nu}),
\label{gt}
\ee
where $\calg_0$ is the $T=0$ conductance scaling function 
given in Eq.~(\ref{gel}),
and $f(E)$ is the Fermi-Dirac distribution function
\eq
f(E)={1\over e^{\beta(E-\delta)}+1}.
\ee
Eq.~(\ref{gt}) is a convolution of the derivative
of the Fermi function (which has width $k_{\rm B}T$)
with the $T=0$ conductance scaling function (which has width $L^{-1/\nu}$), as illustrated in Fig.~2.
In the limit $L\to\infty$, Eqs.~(\ref{g0}) and (\ref{gt}) 
imply that
\eq
\sigma_{xx}(\delta, T)=0
\label{seq0}
\ee
for any $\delta$ if $T\not= 0$:
within the noninteracting theory, the conductivity 
vanishes for all values of the Fermi energy at finite temperature.
This strange result follows from the fact that the set of conducting states is
of measure zero for this transition.  
\begin{figure}   
\center   
\centerline{\epsfysize=1.8in   
\epsfbox{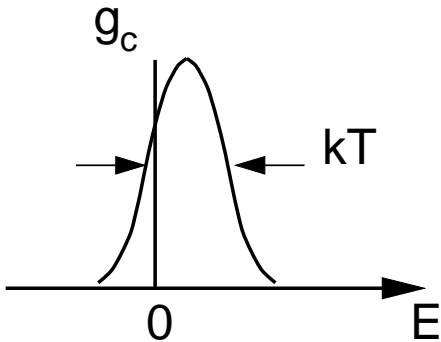}}   
\begin{minipage}[t]{8.1cm}    
\caption{
The convolution of ($-\partial f/\partial E$) with
$\calg_0$ in the thermodynamic limit leads to a vanishing
conductivity at finite temperature in the noninteracting theory.
}  
\label{fig2}     
\end{minipage}    
\end{figure}   

\section{Short-Ranged Interaction, $u\neq0$}

For the conductivity to be non-zero at finite temperatures
near the transition, interactions are necessary, and we now examine the effect of short-range interactions.
Since $u$ is an irrelevant coupling in the RG sense, 
the transitions at $T=0$
are described by the non-interacting fixed point.
In general at such a fixed point, provided the density of states is finite, $z=d$ in $d$-dimensions, and so for the IQHT $z=2$.
Under a RG length scale transformation $b$, $u$ transforms according to
$u^\prime=b^{-\alpha}u$, and energy scales, $\epsilon$, transform as
$\epsilon^\prime=b^z\epsilon$.

\subsection{Naive scaling at $\delta=0$}

The finite temperature conductivity at criticality
is expected to have the scaling form
\eq
\sigma_{xx}(T,u)=b^{2-d}\calg^\prime(b^z T, b^{-\alpha}u).
\label{cscaling0}
\ee
Choosing the scale factor $b=T^{-1/z}$, we obtain a new scaling function
\eq
\sigma_{xx}(T,u)=\calg\left(uT^{\alpha/z}\right).
\label{cscaling1}
\ee
Eq.~(\ref{seq0}) implies, setting $u=0$, that
$\calg(X=0)=0$.

If $u$ were a conventional irrelevant scaling variable
$\calg$ would have a 
power series expansion and one could write
\eq
\sigma_{xx}(T,u)=\calg(0)+\sum_{l=1}^{\infty}(uT^{\alpha/z})^l\calg_l(0).
\label{taylor}
\ee
Since $\calg(0)=0$, Eq.~(\ref{taylor}) implies that
$\sigma_{xx}(T\to0,u)=0$. This result would, paradoxically, exclude the existence
of a conducting critical state.
In fact, as we show in the following sections, 
$\calg(X)$ is a discontinuous function of its argument, $X$, at $X=0$ so that
\eqa
\sigma_{xx}(T\neq0,u=0)&=&\calg(X=0)=0 
\label{singular1}\\
\sigma_{xx}(T\to0,u\neq0)&=&\calg(X\to0)=g_c.
\label{singular2}
\eea
This discontinuous behavior is shown schematically in Fig.~3.
\begin{figure}   
\center   
\centerline{\epsfysize=1.8in   
\epsfbox{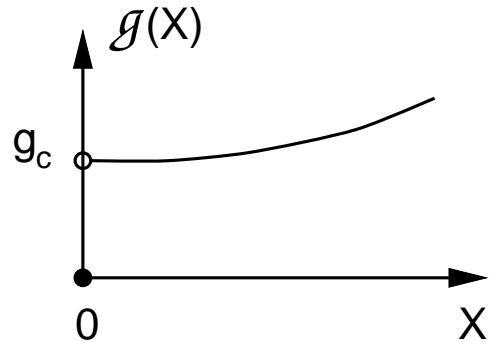}}   
\begin{minipage}[t]{8.1cm}    
\caption{
Discontinuity of the scaling function $\calg(X)$, Eq.~(\ref{cscaling1}), at $X=0$.
}  
\label{fig3}     
\end{minipage}    
\end{figure}

\subsection{Dephasing in the Critical Regime by
Interactions}

For $T\neq0$, interactions, relevant or irrelevant in the RG sense,
will cause transitions between single particle states.\cite{lee-discussion} 
This leads to a finite quasiparticle dephasing 
rate\cite{dephasingfootnote} $\tphi= T^{-p}$.
At a quantum phase transition, the exponent
$p$ that enters the dephasing rate should not be taken from those
for simple disordered metals in the large conductance regime, for it is the 
decay time of the critical eigenstates that matters. This should 
be determined by the underlying critical phenomena. A natural scaling form for the dephasing rate is
\eq
{h\over\tphi}=TY^\prime(b^z T, b^{-\alpha}u),
\label{tauphiprime}
\ee
where the prefactor $T$ is determined by the engineering dimension of $1/\tphi$.
Setting $b=T^{-1/z}$, we have
\eq
{h\over\tphi}=TY(uT^{\alpha/z}).
\label{tauphiscaling}
\ee
As $u$ is an irrelevant coupling (perturbation) which scales towards zero
under renormalization group scale transformations, the unperturbated state 
(noninteracting fixed point) is therefore analytically connected to the 
perturbed state in the presence of $u$. Thus, a perturbative expansion in 
$u$ is justified. To lowest order, $1/\tphi\sim u^2$ from a Fermi's Golden Rule
estimate of the inelastic scattering rate.
Thus, the expected leading scaling behavior 
is
\eq
{1\over\tphi}\sim u_{\rm eff}^2 T \sim u^2T^{1+2\alpha/z},
\ee
or
\eq
\tphi\sim T^{-p},\quad p=1+{2\alpha\over z}.
\label{tauphi}
\ee
For the case of a quantum Hall transition in the presence of
a screening gate, we have $z=2$ and $\alpha\simeq0.65$, and we obtain 
$p\simeq1.65$.

\subsection{Dephasing Length and Thermal Exponent $z_T$}

For a conventional quantum phase transition (with finite
interaction strength at the fixed point), there is one length
scale ($\xi\sim\delta^{-\nu}$) and one time scale ($\Omega^{-1}
\sim\xi^{z}\sim\delta^{-z\nu}$) away from criticality. There are
no finite correlation length or time scales at criticality.\cite{steve}
In such a critical system at finite temperature $T$, one expects to
have one characteristic time $\hbar/T$, the significance of which
is particularly clear in imaginary time, where it sets a finite size in the time direction, as shown in Fig.~4.
However in the present case,
we have obtained an additional (real) time, $\tphi$, which
is much larger than $\hbar/T$ 
as $T\to0$, provided $p>1$ ($\alpha>0$), which is the case if interactions
are irrelevant.  For further discussion of quantum critical transport in 
the incoherent long time limit see Ref.[\onlinecite{subirbook}].

We now turn to the dephasing length, $\lphi$, associated with $\tphi$.
The irrelevance of the interaction at the NIFP allows us to view
the system in terms of weakly interacting diffusive quasiparticles.
The dephasing length that cuts off the phase coherent d.c. transport is
thus
\eq
\lphi=\sqrt{D\tphi}\sim T^{-p/2},
\label{lphi}
\ee 
where $D$ is the diffusion constant
at the non-interacting critical point, obtained from
the wavevector, $q$, and frequency, $\omega$, dependent coefficient, $D(q,\omega)$, in the limit: first 
$q\to0$ and then $\omega\to0$. Thus, anomalous diffusion\cite{john} present
in the opposite limit will not enter our discussion. We show below
that, even though $u$ is irrelevant in the RG sense,
the important length scale introduced by temperature is $\lphi$, so that
\eqa
L_\varphi&\sim& T^{-1/z_T},
\label{lt} \\
z_T&=&{2\over p}={2z\over z+2\alpha}.
\label{zt}
\eea
This length enters the scaling of the transition width in Eq.~(\ref{dbscaling}).
For the IQHT in the presence of short-range interactions,
we thus obtain $z_T\simeq 1.21$.
\begin{figure}   
\center   
\centerline{\epsfysize=2.4in   
\epsfbox{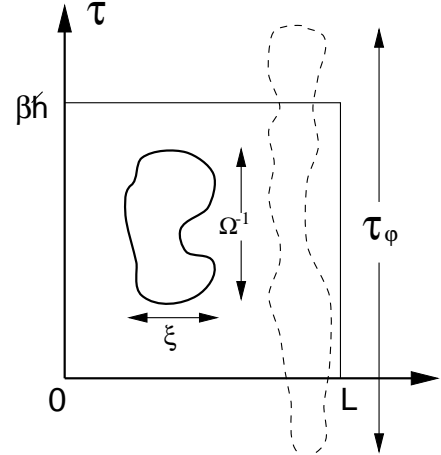}}   
\begin{minipage}[t]{8.1cm}    
\caption{
Schematics of the time and length scales close to a quantum
phase transition. The correlation volume (in space and time) is indicated by the full line for an interacting fixed point. The corresponding volume is indicated by a dashed line in the case where interactions are irrelevant and a
(real) coherence time, $\tphi \gg \hbar/T$, emerges.}  
\label{fig4}     
\end{minipage}    
\end{figure}

\section{Temperature Scaling of Conductivity Near Criticality}

To calculate the conductivity in the presence of a finite dephasing
length, we follow the standard procedure and divide the system into 
$\lphi\times\lphi$ phase coherent blocks. Transport within each block can be described by phase coherent 
single-electron transport using the underlying noninteracting theory.
The disorder-averaged conductivity that we are interested in can be obtained
by averaging over the phase coherent blocks.
The outcome of this exercise is that the system size $L$ in 
Eq.~(\ref{gt}) should be replaced by $\lphi$, which leads to
\eq
\sigma_{xx}(\delta,T,u)=\int dE\left(-{\partial f\over\partial E}\right)
\calg_0(E \lphi^{1/\nu}),
\label{gtphi}
\ee
where $\calg_0$ is a scaling function. Although the precise 
phase coherent geometry appropriate for this averaging procedure 
is unclear, this scaling function is expected to 
have the same {\it qualitative} behavior as $\calg_0$ in Eq.~(\ref{gt}).
Note that this discussion omits contributions to transport
from variable range hopping, which will in fact dominate when $\calg_0$
is very small.

Let $x=\beta(E-\delta)$. We then have
\eq
\sigma_{xx}(\delta,T,u)=-\int dx{\partial f(x)\over\partial x}
\calg_0(x\cdot k_{\rm B}T \lphi^{1/\nu}+\delta\lphi^{1/\nu}),
\label{gtphix}
\ee
where $f(x)=1/(e^x+1)$.

\subsection{At Criticality: $\delta=0$, $T\to0$}

We first study the behavior of the critical conductivity at low temperatures.
At $\delta=0$, the second term in the argument of $\calg_0$ in
Eq.~(\ref{gtphix}) vanishes, leading to
\eq
\sigma_{xx}(\delta=0,T,u)=-\int dx{\partial f(x)\over\partial x}
\calg_0\left[x\cdot \left({T/T_0}\right)^{1-p/2\nu}\right],
\label{gtphix0}
\ee
where $T_0\sim (u^2/D)^{1/(2\nu-p)}$ is a constant determined by
the bare interaction strength and the diffusion constant.
To understand the behavior of $\sigma_{xx}$ which results from Eq.~(\ref{gtphix0}),
one should compare the width of the thermal window, determined by
$-(\partial f/\partial x)$, with the width of the window over which electrons are mobile, determined
by the scaling function ${\cal G}_0(X)$ (see Fig.~1a). There are two different low-$T$ behaviors for $\sigma_{xx}$, depending on the 
value of $p/2\nu$.

\subsubsection{$p<2\nu$: the case of IQHT}

For $p<2\nu$, the argument of the scaling function in Eq.~(\ref{gtphix0})
approaches zero as $T\to0$. Thus, using Eq.~(\ref{gel0}), we have
\eq
\sigma_{xx}(\delta=0,T\to0,u)\simeq \calg_0(X\to0)= g_c.
\label{psmall}
\ee
In this case, the low-$T$ conductance is finite (cf. Eq.~(\ref{singular2}))
(despite the fact that the set of conducting states is of measure zero)
and has a value comparable to the critical phase-coherent conductance 
in the noninteracting theory. 
Hence interactions control the low-temperature behavior, even though they are irrelevant in the RG sense.
The quantum Hall transition with short-range interactions produced
by a screening gate falls into this category since
$p\simeq1.65$ and $\nu\simeq 2.33$ so that $p/2\nu\simeq0.35$.

\subsubsection{$p>2\nu$}

For sufficiently irrelevant interactions (large $\alpha$), the condition
$p>2\nu$ may be satisfied. In this case, the argument of $\calg_0$
in Eq.~(\ref{gtphix0}) diverges as $T\to0$ for fixed $x$. Taking
$\calg_0(X)$ from Eq.~(\ref{gel0}),
\eqa
\sigma_{xx}(\delta=0,T,u)&\simeq&\int dx \calg_0[x\cdot (T_0/T)^{p/2\nu-1}]
\nonumber \\
&\sim& T^{p/2\nu-1}.
\label{plarge}
\eea
Thus the critical conductivity
vanishes as $T\to0$ according to a universal
power law. Note that the power law exponent {\it cannot} be obtained using naive
scaling with irrelevant couplings by following the approach discussed in section III.A.
Again, this vanishes because the set of conducting states is of measure zero.  
The difference between the results for the two cases $p<2\nu$ and $p>2\nu$
will be further elucidated below.

\subsection{Transition Width:
$\delta\neq0$, $T\neq0$}

Hereafter, we specialize to $p<2\nu$ (case 1 above) which is appropriate
for the quantum Hall transition with short-range interactions. For $\delta\neq0$ and small $T$, the first term
in the argument of the $\calg$ in Eq.~(\ref{gtphix}) can be
ignored, leading to
\eq
\sigma_{xx}(\delta,T)\simeq \calg_0(\delta\lphi^{1/\nu}).
\label{sigmatlphi}
\ee
Making use of $\lphi\sim T^{-p/2}=T^{-1/z_T}$ from Eqs.~(\ref{zt})
and (\ref{lphi}), this can be rewritten as,
\eq
\sigma_{xx}(\delta,T)=\calg_0\left({c\delta\over T^{1/z_T\nu}}\right).
\label{sigmat}
\ee
The transition width is determined by the 
value of $\delta$ at which the scaling variable in Eq.~(\ref{sigmat})
is of order one. We obtain
\eq
\delta^*\sim T^{1/z_T\nu}.
\label{deltastar}
\ee
We can view $\delta^*$ as the width of the energy window of states whose localization
length exceeds the phase coherence length.  If the width of this window
exceeds the energy window defined by the Fermi function through the
temperature (i.e., if $z_T\nu > 1$ or equivalently $p<2\nu$), then
the conductivity will scale to a finite value
as discussed above.  Conversely, if the energy window of states is 
narrower than
the temperature, the conductivity becomes sensitive to the fact that the set
of conducting states is of measure zero.

At large argument, the scaling function in Eq.~(\ref{sigmatlphi}) falls
off exponentially with $\lphi/\xi$, being controlled by the crossover to
the non-interacting localized phase. But the interaction $u$, although
irrelevant at the critical fixed point, will give rise to conduction by
variable range hopping in the localized phase. Because $u$ is dangerously
irrelevant in this sense, variable range hopping will not be part of the
universal crossover scaling function in Eq.~(\ref{sigmatlphi}), but
will only set in when $\lphi$ exceeds the hopping length $R_{hop}$.
Naive scaling suggests that the ratio of this longer crossover length to 
$\xi$ will diverge as a power in $\xi$.

\null From Eq.~(\ref{deltastar}) we deduce the temperature scaling exponent $\kappa$ 
for the case of short-range interactions,
\eq
\kappa={1\over z_T\nu}\simeq 0.36.
\label{kappa}
\ee
Interestingly, because the value of $z_T$ happens to be close to 1 - 
the expected value with long-range Coulomb interactions - the value
of $\kappa$ is quite close to the corresponding value
$\kappa\simeq 0.42$ as well, provided that $\nu$ is indeed the same in both cases.
This suggests that temperature scaling of the transition width will not be
dramatically altered by the presence of a screening gate and careful
measurements will need to be made to see the change in the exponent.
An important feature of Eq.~(\ref{sigmatlphi}) is that it implies
that the correct thermal scaling variable is
\eq
{\lphi\over\xi} \quad{\longleftrightarrow}\quad {1\over T\xi^{z_T}},
\label{thermalvariable}
\ee
and the thermal scaling function has the form
\eq
\sigma_{xx}(\delta,T)=\calg_0([T\xi^{z_T}]^{-1/z_T\nu}).
\label{thermalscaling}
\ee
These results suggest that by choosing appropriate scaling
variables, the conductivity can be expressed in terms of a
scaling function which is free of singularities in the limit
of small scaling arguments. This will allow a description
of transport within the conventional scaling framework, 
despite the fact that the scaling function $\calg(X)$ of Eq.~(\ref{cscaling1}) is discontinuous.

\subsection{Conventional Scaling Framework}

The basic scaling form at the noninteracting fixed point reads
\eq
\sigma_{xx}(\delta,T,u)=\calg^\prime(b^{1/\nu}\delta,b^{z} T, b^{-\alpha}u).
\label{gens}
\ee
At scale $b=\xi$, one writes
\eq
\sigma_{xx}(\delta,T,u)=\calg(T\xi^z,u\xi^{-\alpha}),
\label{gprime}
\ee
where, as we have shown earlier, the scaling function has a discontinuity 
when its second argument approaches zero.
In view of Eqs.~(\ref{thermalvariable}) and (\ref{thermalscaling}),
it is convenient to change the scaling variables according to
\eq
(T\xi^z,u\xi^{-\alpha})\longrightarrow (\lphi/\xi,u\xi^{-\alpha}).
\label{change}
\ee
This is possible because
\eq
{\lphi\over\xi}={1\over (T\xi^z)^{p/2}(u\xi^{-\alpha})}.
\label{lphioverxi}
\ee
Hence, we can write as an alternative to Eq.~(\ref{gprime})
\eq
\sigma_{xx}(\delta,T,u)=\calgr(\lphi/\xi,u\xi^{-\alpha}),
\label{regular}
\ee
in which $\calgr$ is a regular scaling function when its second
argument is taken to zero. Specifically,
\eq
\calgr(\lphi/\xi,0)=\calgr(\delta^\nu/T^{1/\zt},0)
=\calg_0(\delta/T^{1/\zt\nu}),
\ee
where use has been made of Eq.~(\ref{sigmatlphi}) in
the last step and the behavior of $\calg_0(X)$ is shown in Fig.1a.
It is perhaps important to note that the change of variables in
Eq.~(\ref{change}) has not removed the singularity associated
with the scaling function in Eq.~(\ref{gprime}). Instead, it simply
makes the singularity inaccessible in 
Eq.~(\ref{regular}), since $u\to0$ implies $\lphi\to\infty$.

\section{Frequency Scaling At $T=0$}

\subsection{Noninteracting case, $u=0$}

For studying the frequency scaling,
we start by returning to the noninteracting theory.\cite{sondhi-discussion}
Scaling implies
\eq
\sigma_{xx}(\delta,\omega)=\calg_0^\prime(b^{1/\nu}\delta,
b^z\omega).
\ee
Putting $b=\xi$ leads to
\eq
\sigma_{xx}(\delta,\omega)=\calg_0(\omega\xi^z).
\label{g0omega}
\ee
The behavior of the scaling function in Eq.~(\ref{g0omega})
is expected from the Mott formula to be
\eq
\calg_0(X)=\left\{ \begin{array}{ll}
X^2\ln^{d-1}X, & X\to0, \\
{\rm const}, & X\to\infty, \end{array} \right.
\label{g0omegax}
\ee
and has been studied numerically.\cite{gammel2}
Thus the natural frequency scaling variable is $\omega\xi^z$, in contrast
to the temperature scaling variable, $T\xi^\zt$, which appears in 
Eqs.~(\ref{thermalscaling}) and (\ref{gens}).

\subsection{Short-range Interactions, $u\neq0$}

Including $u$ as in Eq.~(\ref{gprime}), we write
\eq
\sigma_{xx}(\delta,\omega)=\calg(\omega\xi^z,u\xi^{-\alpha}).
\ee
This function has a nonsingular limit, {\it i.e.}
$\calg(X,Y\to0)=\calg_0(X)$. Thus we conclude that
frequency scaling is conventional, so long as 
$p<2\nu$. Anticipating that this is the case for the IQHT with short-range interactions, further
subtleties that occur in the opposite limit ($p>2\nu$) will not
be discussed here. The transition width
for $\omega\neq0$ but $T=0$ is determined by setting 
$\omega\xi^z(\delta^*)=1$, giving
\eq
\delta^*(T=0,\omega)\sim \omega^{1/ z\nu}.
\label{deltastaromega}
\ee
This should be contrasted with
$\delta^*(T,\omega=0)\sim T^{1/\zt\nu}$ where
$\zt=2/p$, Eq.~(\ref{deltastar}).

\subsection{Irrelevance of Frequency Dephasing}

A finite frequency can also lead to dephasing through interactions.
For $u=0$, the only length scale introduced by a finite
frequency is 
\eq
L_\omega=\sqrt{D/\omega}.
\label{lomega}
\ee
However, when $u\neq0$,
there is a frequency-induced dephasing time $\tphi(\omega)$ which
can be accounted for by including $\omega$ in the discussion of 
section III.B. Following Eqs.~(\ref{tauphiprime}-\ref{tauphi}),
one obtains,
\eq
{1\over\tphi(\omega)}\sim u^2\omega^p
\ee
at $T=0$. This leads to another frequency-dependent
length scale in the diffusive regime,
$L_\omega^u=\sqrt{D\tphi(\omega)}$. Thus
\eq
L_\omega^u\sim\sqrt{D/u^2}\omega^{-1/\zt}.
\ee
The ratio of the two lengths is
\eq
{L_\omega^u\over L_\omega}\sim \omega^{-(z-\zt)/z\zt}.
\ee
Provided interactions are irrelevant, so that $\alpha>0$ and $\zt<2$ from
Eq.~(\ref{zt}), this ratio diverges in the limit $\omega\to0$.
The fact that $L_\omega^u\gg L_\omega$ ensures that frequency
dephasing results only in corrections to scaling of the
conductivity, and is irrelevant in the asymptotic limit.

\section{General Temperature and Frequency Scaling}

In this section, we discuss the general scaling behavior of the conductivity
as a function of both frequency and temperature. We start with the basic
scaling form at the NIFP,
\eq
\sigma_{xx}(\delta,T,\omega,u)=\calg(T\xi^z,\omega\xi^z,u\xi^{-\alpha}).
\label{gallz}
\ee
We convert to new scaling variables as in Eq.~(\ref{lphioverxi}). Then
\eq
\sigma_{xx}(\delta,T,\omega,u)=\calgr(\lphi/\xi,\omega\xi^z,u\xi^{-\alpha}),
\label{gallzt}
\ee
where $\calgr(X,Y,Z)$ is continuous in $Z$ at $Z=0$. Let 
\eq
\calgr(\lphi/\xi,\omega \xi^z,0)=\calg_0(\lphi/\xi,\omega \xi^z).
\label{a}
\ee
Thus for $\xi\gg1$ we have
\eq
\sigma_{xx}(\delta,T,\omega,u)=\cala(T\xi^\zt,\omega\xi^z).
\label{gazt}
\ee
Now consider the approach to the critical point at $\delta=0$. As $\xi\to\infty$,
one argument of $\calg_0$ diverges and the other approaches zero, but the scaling variable
\eq
{(\lphi/\xi)^z\omega\xi^z}={\omega\over T^p},
\label{finitevar}
\ee
remains finite for $\omega,T\neq0$. (We have used $\zt=2/p$ and $z=2$.) 
Thus at the critical point
\eq
\sigma_{xx}(\delta=0,T,\omega,u)=\cala\left(\omega\tphi\right)
=\cala\left({\omega\over T^p}\right).
\label{gafinal}
\ee
We see that $\omega/T^p \sim \omega/T^{1.65}$  
is the scaling variable at criticality,
in contrast to the conventional situation in which
the interaction $u$ scales to a finite value at the fixed point and the scaling variable is $\omega/T$.

\section{Summary}

We have shown that, in the presence of short-range Coulomb interactions,
the integer quantum Hall transition is a quantum phase
transition of an unconventional kind.
We find that
the interactions, though irrelevant, are responsible for the existence of a finite critical 
conductivity.
In addition, the
conventional $\omega/T$ scaling at criticality is replaced by
$\omega/T^p$ scaling, where $p$ is a critical exponent
controlling the inelastic dephasing time. 
As a result, there exist two independent dynamical scaling
exponents $\zt\neq z$ for temperature and frequency respectively.
The dynamic exponents determine the physical length scales associated with 
$T$ and $\omega$: $(L_\varphi,L_\omega)\sim(T^{-1/\zt},T^{-1/z})$.
These unconventional results follow from the fact that, though short-range
interactions are irrelevant at the critical point, 
the physical behavior is discontinuous in the interaction strength in the non-interacting limit.
Associated with this is the existence of a 
coherence time much longer than the conventional quantum coherence time, 
$\hbar/T$,  as interactions scale to zero and the system
scales towards the non-interacting fixed point. We have shown that the scaling exponent $\zt$ (or $p$) is
completely determined by the scaling dimension of the leading irrelevant
interaction. 
The physics discussed here may in fact be quite general for quantum
critical transport phenomena such as the conventional 
Anderson-Mott metal-insulator transitions, 
whenever the interactions scale to zero at the fixed point. 

For the IQHT with short-range interactions, we have the set of critical exponents
\eq
\nu\simeq2.3,\quad \zt\simeq1.2, \quad z=2,
\label{exponents}
\ee
which describe the scaling with sample size, temperature and
frequency according to Eq.~(\ref{dbscaling}).

This behavior can be checked experimentally, for example by looking for a
change in the temperature scaling of the transition width whose exponent
will change from $\kappa \simeq 0.42$ to $\kappa \simeq 0.36$, or by looking
at the frequency/temperature scaling described in Eq.~(\ref{gafinal}) where a
larger change in exponent is expected.
The experimental requirement is that
the long-range Coulomb interaction between electrons at large
distances be screened, so that they interact via a residual, short-range
interacting potential.

\vspace{1.0truecm}

\centerline{\bf ACKNOWLEDGMENTS}
\bigskip
We would like to thank Assa Auerbach, Bodo Huckestein, Subir Sachdev, 
and especially Dung-Hai Lee and Shivaji Sondhi for many valuable 
discussions. JTC is supported by EPSRC Grant No. GR/MO4426. MPAF 
is supported by NSF Grants DMR-97-04005 and DMR-95-28578.
SMG is supported by NSF Grant DMR-9714055. ZW is supported by
DOE Grant DE-FG02-99ER45747 and an award from Research Corporation.
The authors would like to 
thank the Institute for Theoretical Physics at UCSB where this work was begun
and the generous support of NSF Grant PHY94-07194.


\begin{references}
\bibitem{notelambda} A short-distance cut-off can be introduced in 
Eq.~(\ref{h}) for the case of $\lambda\ge4$.
\bibitem{dephasingfootnote}
Note that one should distinguish
between the electron-electron inelastic scattering time $\tee$, which is
a single particle (energy) relaxation time, and the two-particle
or phase coherence time $\tphi$ defined in the weak localization theory.\cite{boris}
The differences between
the two in a disordered metal depends on dimensionality, although on 
general grounds $\tphi \leq \tee$. It turns out within perturbative
weak localization theory \cite{boris},
$\tphi^{-1}\sim\tee^{-1}\sim T^{3/2}$ in 3d, $\tphi^{-1}\sim T^{2/3}$
and $\tee^{-1}\sim T^{1/2}$ in 1d. In the marginal case of $d=2$,
$\tphi\sim \tee\sim T^{-1}$ up to possible $\log T$ corrections.
Here $\tphi$ can also be formally defined in terms of the temperature
dependence of the conductivity (although it is not weak localization).
The result for the temperature dependence here is different from the weak localization 
case and contains the exponent $\alpha$
because the disorder eigenstates are those for a critical point, not a metallic
phase.  We have assumed that in perturbation theory for an irrelevant interaction, both
$\tee$ and $\tphi$ scale with $|u_{\rm eff}|^2$.
\end{references}
\end{document}